\documentclass[twocolumn,tighten,trackchanges,twocolappendix]{aastex63}

\usepackage{amsmath}
\usepackage[normalem]{ulem}

\newcommand{\swift}{{\rm Swift}}

\newcommand{\oiiiblong}{\hbox{[O\sc iii]\,$\lambda$5007}}

\newcommand{\oiii}{\hbox{[O\sc iii]}}

\newcommand{\ha}{H$\alpha$}
\newcommand{\hb}{H$\beta$}

\newcommand{\oivlong}{\hbox{[O\textsc{iv}]\,25.9\,\micron}}

\newcommand{\msun}{$M_{\odot}$}

\newcommand{\lsun}{$L_{\odot}$}
\newcommand{\kms}{km\,s$^{-1}$}

\newcommand{\lumcgs}{erg\,s$^{-1}$}

\newcommand{\lumx}{$L_{\rm 2-10\,keV}$}
\newcommand{\lummir}{$L_{\rm 5-38\,\mu m}$}
\newcommand{\sfrunit}{$M_{\odot}$\,yr$^{-1}$}

\newcommand{\pscm}{cm$^{-2}$}
\newcommand{\nh}{$N_{\rm H}$}

\received{October 30, 2020}
\revised{November 12, 2020}
\accepted{November 13, 2020} %\today
\submitjournal{ApJL}

\shorttitle{Faint AGN in ULIRG with Powerful Outflow}
\shortauthors{Chen et al.}
\begin{document}

\title{NuSTAR Non-detection of a Faint Active Galactic Nucleus in an Ultraluminous IR Galaxy with Kpc-scale Fast Wind}

\correspondingauthor{Xiaoyang Chen}
\email{xiaoyang.chen@nao.ac.jp}

\author[0000-0003-2682-473X]{Xiaoyang Chen}
\affiliation{National Astronomical Observatory of Japan, 2-21-1, Osawa, Mitaka, Tokyo 181-8588, Japan}

\author[0000-0002-4377-903X]{Kohei Ichikawa}
\affiliation{Astronomical Institute, Tohoku University, 6-3 Aramaki, Aoba-ku, Sendai, Miyagi 980-8578, Japan}
\affiliation{Frontier Research Institute for Interdisciplinary Sciences, Tohoku University, 6-3 Aramaki, Aoba-ku, Sendai, Miyagi 980-8578, Japan}

\author[0000-0001-6020-517X]{Hirofumi Noda}
\affiliation{Department of Earth and Space Science, Graduate School of Science, Osaka University, 1-1 Machikaneyama-cho, Toyonaka-shi, Osaka 560-0043, Japan}
\affiliation{Project Research Center for Fundamental Sciences, Osaka University, 1-1 Machikaneyama, Toyonaka, Osaka 560-0043, Japan}

\author[0000-0002-6808-2052]{Taiki Kawamuro}
\affiliation{National Astronomical Observatory of Japan, 2-21-1, Osawa, Mitaka, Tokyo 181-8588, Japan}
\affiliation{Nu\'{c}leo de Astronom\'{i}a de la Facultad de Ingenier\'{i}a, Universidad Diego Portales, Av. Ej\'{e}ercito Libertador 441, Santiago, Chile}

\author[0000-0002-3866-9645]{Toshihiro Kawaguchi}
\affiliation{Department of Economics, Management and Information Science, Onomichi City University, Hisayamada 1600-2, Onomichi, Hiroshima 722-8506, Japan}

\author[0000-0002-3531-7863]{Yoshiki Toba}
\affiliation{Department of Astronomy, Kyoto University, Kitashirakawa-Oiwake-cho, Sakyo-ku, Kyoto 606-8502, Japan}
\affiliation{Academia Sinica Institute of Astronomy and Astrophysics, 11F of Astronomy-Mathematics Building, AS/NTU, No.1, Section 4, Roosevelt Road, Taipei 10617, Taiwan}
\affiliation{Research Center for Space and Cosmic Evolution, Ehime University, 2-5 Bunkyo-cho, Matsuyama, Ehime 790-8577, Japan}

\author[0000-0002-2651-1701]{Masayuki Akiyama}
\affiliation{Astronomical Institute, Tohoku University, 6-3 Aramaki, Aoba-ku, Sendai, Miyagi 980-8578, Japan}



\begin{abstract}

Large-scale outflows are generally considered as a possible evidence that active galactic nuclei (AGNs) can severely affect their host galaxies. 
Recently an ultraluminous IR galaxy (ULIRG) at $z$\,=\,0.49, AKARI J0916248+073034, 
was found to have a galaxy-scale \oiiiblong\ outflow with one of the highest energy-ejection rates at $z$\,$<$\,1.6. 
However, the central AGN activity estimated from its torus mid-IR (MIR) radiation is weak relative to the luminous \oiii\ emission. 
In this work we report the first NuSTAR hard X-ray follow-up of this ULIRG to constrain its current AGN luminosity. 
The intrinsic 2--10 keV luminosity shows a 90\% upper-limit of $3.0\times10^{43}$\,\lumcgs\ 
assuming Compton-thick obscuration (\nh\,=\,$1.5\times10^{24}$\,\pscm),
which is only 3.6\% of the luminosity expected from the extinction corrected \oiii\ luminosity.
With the NuSTAR observation, we succeed to identify that 
this ULIRG has a most extreme case of X-ray deficit among local ULIRGs. 
A possible scenario to explain the drastic declining in both of the corona (X-ray) and torus (MIR) is that 
the primary radiation from the AGN accretion disk is currently in a fading status,  
as a consequence of a powerful nuclear wind suggested by its powerful ionized outflow in the galaxy scale. 
%\sout{The fading process happens in a timescale of $10^4$--$10^6$\,yr, which is much shorter than the duration of starburst in ULIRGs of $\sim$\,$10^8$\,yr, implying that although the AGN can drive an strong wind, it can be decayed in short period, and the net effect of the feedback on stellar build-up in the galaxy can be limited as indicated with the high star formation rate, i.e., 1000\,\sfrunit.}

\end{abstract}

\keywords{Active galaxies -- X-ray AGN -- ULIRG}


%%%%%%%%%%%%%%%%%%%%%%%%%%%%%%%%%%%%%%%%%%%%%%%%%%%%%%%%%%%%%%%%%%%%%%%%%%%%%%%%
%%%%%%%%%%%%%%%%%%%%%%%%%%%%%%%%%%%%%%%%%%%%%%%%%%%%%%%%%%%%%%%%%%%%%%%%%%%%%%%%
%%%%%%%%%%%%%%%%%%%%%%%%%%%%%%%%%%%%%%%%%%%%%%%%%%%%%%%%%%%%%%%%%%%%%%%%%%%%%%%%

\section{Introduction} \label{sec:intro}

Ultraluminous IR galaxies ($L_{\rm IR}$\,$>$\,$10^{12}$\,\lsun, ULIRGs) are a population of galaxies emitting nearly entire energy in IR band 
\citep{Sanders1996}. %Robinson2000 
The high $L_{\rm IR}$ originates from dust heated by UV\,/\,optical radiation of vigorous starbursts and/or active galactic nuclei (AGNs). 
ULIRGs are thought to represent a rapidly growing phase of massive galaxies in the transition from disk to elliptical galaxies, as gas and dust are swept out by starburst- and\,/\,or AGN-induced outflows \citep[e.g.,][]{Hopkins2008}. 
%It is widely accepted that a powerful AGN is necessary to drive outflow exceeding the escape velocity of the host galaxy, e.g., $v$\,$\ge$\,$500$\,\kms \citep[e.g.,][]{Harrison2012}. 
%The study on outflows in ULIRGs is important to understand the feedback effect of the activity of supermassive black holes (SMBH) on the growth of spheroidal components of their host galaxies.

\citet{Chen2020} reported a new follow-up program for sources in the AKARI Far-Infrared Surveyor (FIS) Bright Source Catalogue\footnote{https://www.ir.isas.jaxa.jp/AKARI/Observation/update/\\20160425\_preliminary\_release.html}\,(ver.2)
to construct a statistical flux-limited sample of ULIRGs at intermediate redshifts ($z$\,=\,0.5\,--\,1). 
Among the sources, AKARI J0916248+073034 (hereafter J0916a) at $z_{\rm spec}$\,=\,0.49, indicates signatures of an extremely strong ionized-gas outflow in the Subaru\,/\,FOCAS long-slit spectroscopic image \citep[with slit width of 0.5\arcsec,][]{Chen2019}. 
% associated with a high star formation rate (SFR) of approximately 1000\,\sfrunit 
The \oiiiblong\ emission line has a FWHM of 1830\,\kms\ and a shift of $-770$\,\kms\ relative to stellar absorption lines. 
%Furthermore, low-ionization \oiilong\ doublet also shows a large FWHM of 910\,\kms\ and a velocity shift of $-380$\,\kms. 
The long-slit spectroscopic image shows that the outflow extends to a radius of 4 kpc. 
The mass-loss and energy-ejection ($\dot{E}_{\rm k}$) rates are estimated to be 500\,\sfrunit\ and $10^{44.6}$\,\lumcgs, respectively, 
implying that J0916a has one of the highest $\dot{E}_{\rm k}$ ionized outflows among ULIRGs\,/\,AGNs at $z$\,$<$\,1.6 and it is comparable to the most powerful outflows in quasars at $z$\,$\sim$\,2 \citep[e.g.,][]{Harrison2012, Zakamska2016}.  %, Carniani2015, Bischetti2017
Emission line ratios indicate that the outflow is driven by AGN. 

The strong \oiii\ emission line ($10^{43.8\pm0.5}$\,\lumcgs\ after extinction correction\footnote{
The extinction is estimated to be $E(B-V)$\,=\,$1.0\pm0.3$ using Balmer decrement, which is close to the typical amount of dust attenuation in local ULIRGs \citep[e.g.,][]{Garcia2009}.}) %Alonso2006, 
implies that the AGN is luminous with bolometric luminosity\footnote{Throughout the paper we use $L_{\rm band,\,indicator}$ to denote the luminosity in a given band or wavelength range estimated from the employed indicator.}\,($L_{\rm bol,\,[OIII]}$) of $10^{46.3\pm0.5}$\,\lumcgs\ 
using the empirical relationships \citep{Ueda2015, Ricci2017}.
However, the mid-IR (MIR) radiation originating from dusty torus in the vicinity ($\sim$\,10\,pc) of the SMBH is weak. 
The 5--38\,\micron\ MIR luminosity ($L_{\rm 5-38\,\mu m}$) of J0916a is $10^{44.9\pm0.1}$\,\lumcgs, 
which is integrated using the best-fit SED of AGN component reported in \citet{Chen2019}\footnote{There is a mistake in Table 3 of \citet{Chen2019}. The reported value of $L_{\rm bol}^{\rm AGN}$ was not the bolometric luminosity of AGN, but the total integrated luminosity of the best-fit AGN SED including the torus thermal and scattering radiation and the transmitted primary emission.}.
The corresponding bolometric luminosity with the empirical function \citep{Ichikawa2019s} 
is $L_{\rm bol,\,torus}$\,=\,$10^{45.8\pm0.1}$\,\lumcgs, which is only 32\% of $L_{\rm bol,\,[OIII]}$ 
and would indicate that the central engine of the AGN in J0916a is declining, 
i.e., the AGN is currently less active than its past epoch, 
while the observed strong \oiii\ emission with extreme outflow reflect a historical effect of the AGN during its preceding active phase, 
due to the time-lag between AGN activity in the nuclear region and outflow in the galaxy scale \citep[e.g.,][]{Harrison2017}. 

Recent works reported the luminosity declining of AGN within a timescale of $10^{3}$--$10^{4}$ years in a population called ``fading AGN'' or ``dying AGN'' \citep[e.g.,][]{Schawinski2010}. This population shows AGN signatures in large spatial scales, e.g., radio jets and\,/\,or bright \oiii\ line in the kpc-scale narrow line region (NLR), but lack the features in small scales, e.g., weak or lack of X-ray (corona) and\,/\,or MIR (torus) emission. 
The faintness in small scales of those objects implies a transient stage that the central engine was active in the past, but currently seems quenched \citep{Kawamuro2017, Ichikawa2019a, Ichikawa2019b, Cooke2020}. %Schirmer2013, Ichikawa2016, Villar2018
The ``changing-look AGN'' is another population of AGNs which shows drastically declining by an order of magnitude 
in 1--10 years \citep[e.g.,][]{LaMassa2015, MacLeod2016} as their accretion states change from bright (standard disk) phase to faint (radiatively inefficient accretion flow) phase \citep[e.g.,][]{Noda2018, Ruan2019}. 
Recent X-ray studies of ULIRGs presented that the coronae in ULIRGs could be intrinsically weak compared to normal AGNs \citep[e.g.,][]{Teng2015}.

In order to constrain the current activity of the AGN in J0916a, hard X-ray observation over 10 keV is required with its power to penetrate the heavy obscuration in the nuclear region \citep[e.g.,][]{Ricci2015}. 
There is no previous X-ray observation for J0916a and it is not detected in the \swift/BAT 14-195 keV survey due to its faintness at a relatively high redshift. 
In this letter we report the first hard X-ray follow-up of J0916a with NuSTAR \citep{Harrison2013}. 
Thanks to a wide energy range over 10 keV with great sensitivity, 
NuSTAR makes it capable to directly constrain the intrinsic AGN luminosity in J0916a even in the case with Compton-thick absorption, e.g., gas column density (\nh) of at least $1.5\times10^{24}$\,\pscm. 
%The NuSTAR data reduction and X-ray analysis is reported in Section \ref{sec:data} and we discuss on the results in Section \ref{sec:discussions}. 
Throughout the paper we adopt the cosmological parameters, $H_0=$\,70\,\kms\,Mpc$^{-1}$, $\Omega_{\rm M}=0.3$ and $\Omega_{\Lambda}=0.7$. 

% However, \textbf{the $\bm L_{\mathbf{2-10}}$ estimated from MIR luminosity of the AGN
% is much fainter ($\bm{L}_{\mathbf{2-10}}$\,$\bm =$\,$\mathbf{10^{44.1\pm0.2}}$\,\lumcgs).}\,The $L_{\mathrm{MIR}}$, which originates from dusty torus in the vicinity ($\bm \sim$10\,pc) of the SMBH, 
% is estimated from the SED fitting ($10^{45.2\pm0.1}$\,\lumcgs, Fig.\ref{fig:image_sed}), 

% \textbf{Hence the IR observation indicates that the central AGN in J0916a is declining.}

% \textbf{The extremely strong galaxy-scale outflow associated with a less luminous AGN central engine in J0916a is puzzling}.
% There are two possible scenarios. % that can account for these contradictory results between the larger (> 1 kpc) and smaller (< 10 pc) scales. 
% One explanation is that \textbf{(1) currently the central AGN is less active than its peak epoch (hereafter fading scenario).}
% The low $L_{\rm MIR}$ from the 10 pc-scale torus
% suggests the AGN currently in a fading status; 
% while the observed extreme outflow and strong \oiii\ emission reflect a historical effect of the central engine during its preceding active phase, 
% due to the time-lag between AGN activity in the nuclear region and outflow in the galaxy-scale (Harrison et al.\,2017). 

%%%%%%%%%%%%%%%%%%%%%%%%%%%%%%%%%%%%%%%%%%%%%%%%%%%%%%%%%%%%%%%%%%%%%%%%%%%%%%%%
%%%%%%%%%%%%%%%%%%%%%%%%%%%%%%%%%%%%%%%%%%%%%%%%%%%%%%%%%%%%%%%%%%%%%%%%%%%%%%%%
%%%%%%%%%%%%%%%%%%%%%%%%%%%%%%%%%%%%%%%%%%%%%%%%%%%%%%%%%%%%%%%%%%%%%%%%%%%%%%%%

\begin{figure*}[ht!]
    \begin{center}
	\includegraphics[width=0.32\textwidth]{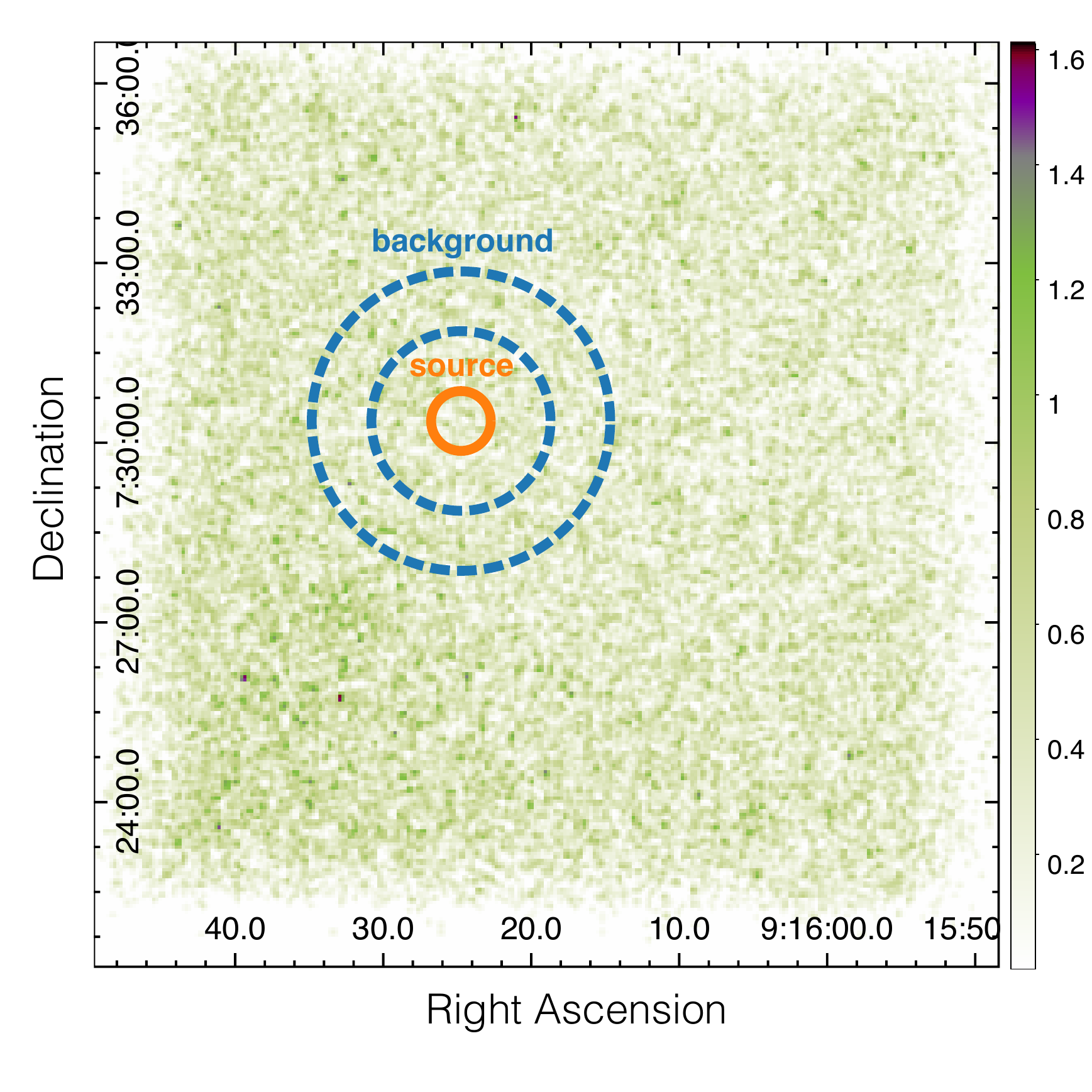}
	\includegraphics[width=0.32\textwidth]{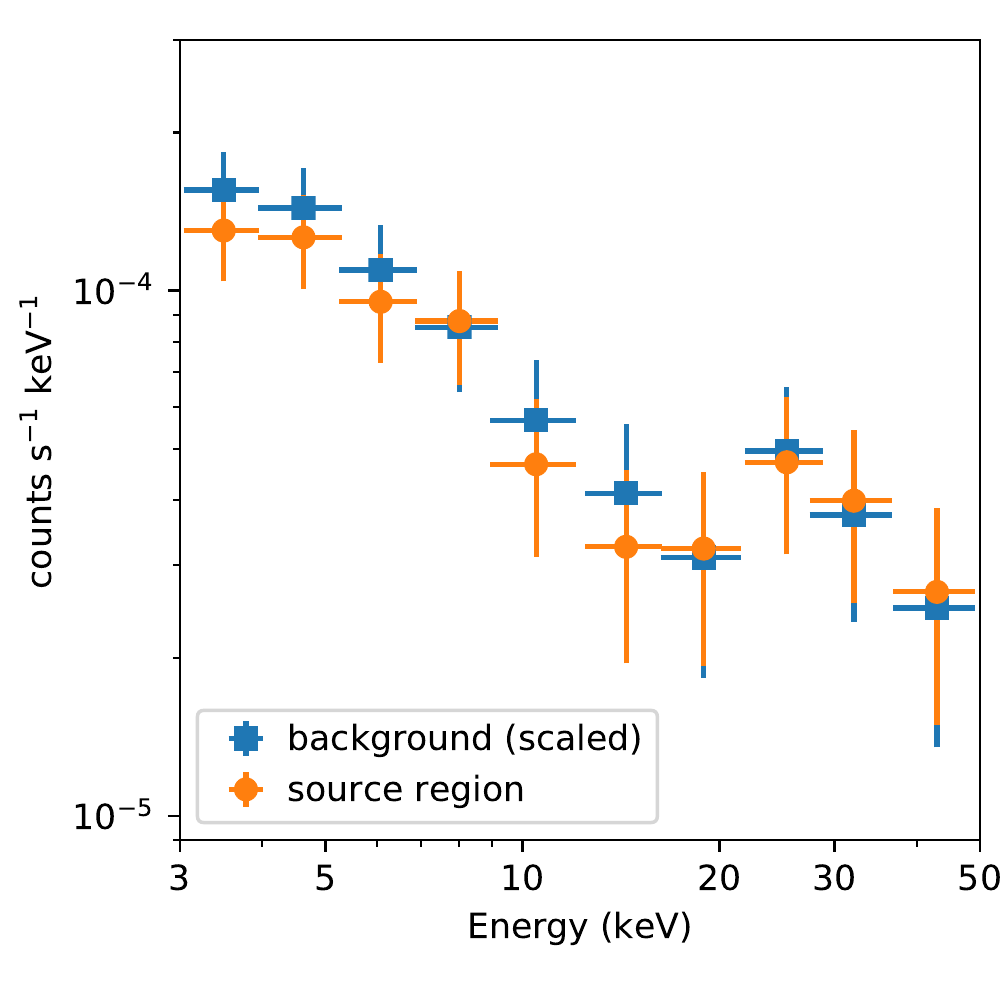}
	\includegraphics[width=0.32\textwidth]{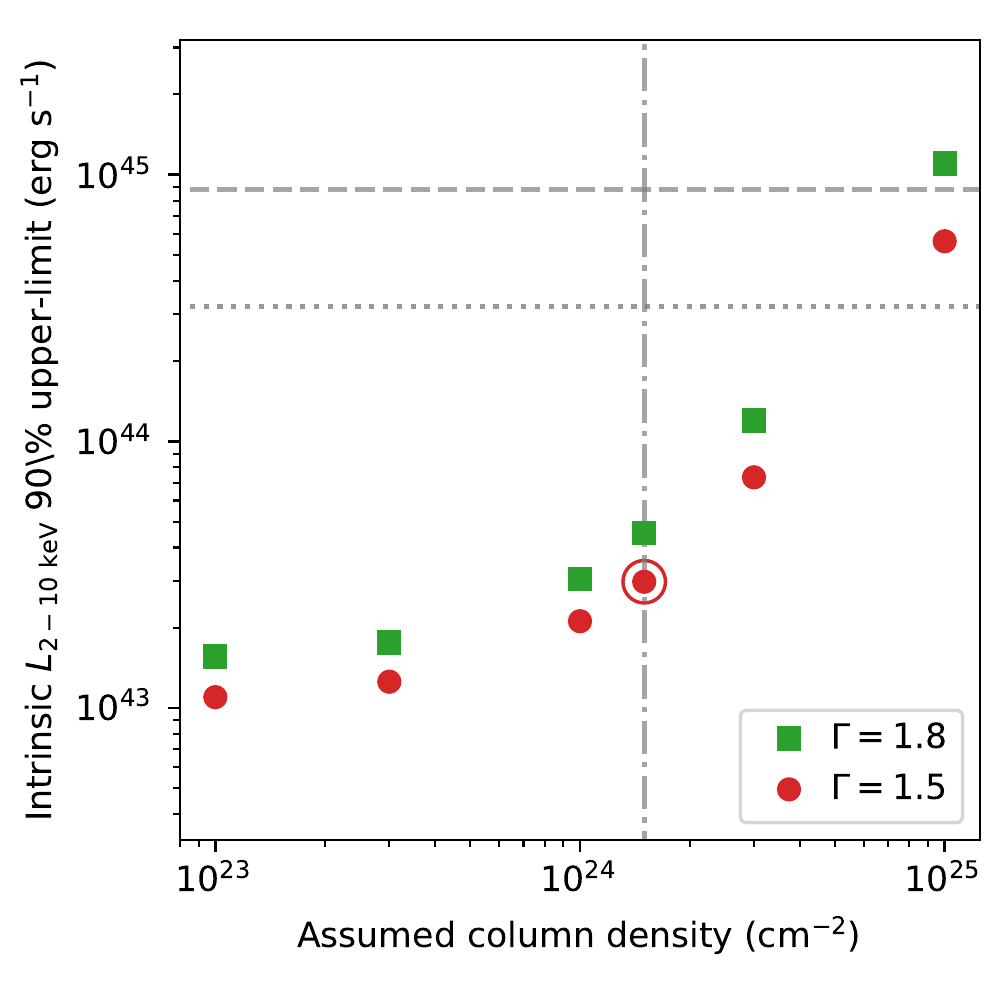}
    \end{center}
    \vspace{-25pt}
	\caption{
	\textbf{Left:} NuSTAR 6--24 keV image in units of counts per pixel. The image is co-added using the FPMA and FPMB data from the two observation epochs, and smoothed using a Gaussian profile with a radius of 2 pixels. 
	The source (circular) and background (annular) regions are shown in orange and blue, respectively. 
	\textbf{Middle:} NuSTAR 3--50 keV spectra extracted from the source (orange) and background (blue) regions shown in the left panel. 
	\textbf{Right:} 
	Estimated 90\% upper-limit of intrinsic \lumx\ with different assumptions of photon index ($\Gamma$) and intrinsic obscuration (\nh). The two horizontal lines denote the \lumx\ converted from \oiiiblong\ (dashed) and 5--38 \micron\ (dotted) luminosities. The vertical dash-dotted line shows the threshold of Compton-thick obscuration. The open circle denotes the fiducial estimation in the discussion.
	}
	\label{fig:iamge_spec}
\end{figure*}

\section{NuSTAR Constraint on the Intrinsic X-ray luminosity of J0916a} \label{sec:data}

J0916a was observed by NuSTAR on 2020 June 16 and 17 with a total on-source exposure of 105.6 ksec (NuSTAR GO cycle-6 program 06108, PI:X.Chen). 
The data was reduced using the NuSTAR data analysis pipeline (\textit{nupipeline}, ver.1.9.2) with the latest HEASARC's calibration database (CALDB, ver.20200720). 
The South Atlantic Anomaly (SAA) filtering is adopted to reduce the observation time intervals affected by the enhancement of background event rates from passage through the SAA and the so-called ``tentacle'' region with the \textit{nupipeline} configuration of \texttt{saacalc=1 saamode=optimized tentacle=yes}. %\footnote{https://www.nustar.caltech.edu/page/background}
The calculation algorithm ``1'' is chosen since it provides the best noise removal as shown in the background filtering reports for both of the two observation epochs\footnote{
The default setting \texttt{saacalc=3 eliminatesource=yes} evaluates count rates after the exclusion of the contribution of the brightest sources in the field of view, which could results in noise remnants for this dataset, because the object is very weak and the brightest sources can be dominated by noise.}.
We use the ``optimized'' mode since it provides comparable noise removal result with the ``strict'' mode but does not lose large amount of exposure as the later one ($\sim10\%$).  
After removal of bad time intervals, the final exposure is 37.3 ksec (96.6\%) and 67.0 ksec (99.8\%) on source for the first and second observation epochs, respectively. 
Totally four cleaned events are created from the data obtained by the two focal plane modules (FPMA and FPMB) during the two epochs. 

The images from the four cleaned events are merged using the FTOOLS package \textit{fimage} and the co-added 6--24 keV image is shown in the left panel of Figure \ref{fig:iamge_spec}. 
No serendipitous sources are detected in the fields of view of the NuSTAR observations. 
A circular region with a radius of 30\arcsec\ centered at the optical position of J0916a is employed as the source region, following the suggestion for faint objects in Section 4.3 of the NuSTAR Data Analysis Software Guide\footnote{
We also checked the spectrum extracted using a larger radius of 60\arcsec\ with EEF of 90\%. 
The net spectrum shows -39.6 counts ($-1.5\sigma$) in 6--24 keV. 
The difference of estimated fluxes is about 5\%, thus having little impact on our conclusion.}.
A radius of 30\arcsec\ corresponds to a fraction of encircled energy (EEF) of 65\%, which is taken into account in the later analysis.
A surrounding annular region with radii of 90\arcsec\ and 150\arcsec\ is used to estimate the background level. 
For each of the four cleaned events the spectra are extracted from the source and background regions.
The co-added source and background (scaled to the source region) spectra  
obtained with the FTOOLS package \textit{addspec} are shown in the middle panel of Figure \ref{fig:iamge_spec}. 
There is no significant difference between the source and background spectra.
The net spectrum (i.e., source\,$-$\,background) shows 
$-9.9$ counts in 3--6 keV, 
$-16.9$ counts in 6--24 keV, and 
$8.5$ counts in 24--50 keV, 
which correspond to $-1.1\sigma$, $-1.3\sigma$, and $0.6\sigma$, respectively, 
where $\sigma$ is estimated from the poisson noise within the source region in each energy range.
Therefore we conclude that the object is not detected. 

Spectral analysis is not feasible as the spectrum is dominated by background. 
In order to estimate the upper-bound of the X-ray luminosity, 
we employed a fixed, putative torus model of \citet{Ikeda2009} following \citet{Ichikawa2019b}, 
which takes account of an torus-absorbed and Compton scattered power-law component, 
a reflected continuum and an accompanying fluorescent iron K$\alpha$ line. 
Only the normalization of the power-law component is set as a free parameter. 
The cut-off energy of the power-law component is fixed to 360 keV. 
The opening angle of the torus is fixed to 60\arcdeg, which reflects a typical covering factor for X-ray selected AGNs \citep[e.g.,][]{Stalevski2016, Ichikawa2019s}. %Mateos2017
An inclination angle of 80\arcdeg\ is selected since J0916a is identified as a Seyfert 2 galaxy from its optical spectrum\footnote{
We have confirmed that the choice of the opening angle (45\arcdeg--70\arcdeg) and inclination angle (70\arcdeg--89\arcdeg) does not significantly affect our conclusion.}. 
We consider two values of photon index ($\Gamma$), i.e., 1.8, which is a typical value of normal Seyfert galaxies \citep[e.g.,][]{Ricci2017}, 
and 1.5, as shown in the harder spectra found in several ULIRGs \citep{Teng2014, Teng2015, Oda2017}.
Galactic absorption of \nh\,=\,$2.6\times10^{20}$\,\pscm, which is given by the FTOOLS package \textit{nh}, 
and 
intrinsic obscuration with \nh\ from $10^{23}$ to $10^{25}$\,\pscm\ are employed in the estimation.  
We estimate the upper-limit of the intrinsic 2--10 keV luminosity (\lumx) by simulating artificial spectra with the above assumed models to achieve net counts of 3$\sigma$ in 6--24 keV range. 
The estimated 3$\sigma$ upper-limits with different assumptions of $\Gamma$ and \nh\ are then converted to 90\% upper-limits considering that the 90\% confidence level corresponds to 1.645$\sigma$ for a normal distribution, and the results are shown in the right panel of Figure \ref{fig:iamge_spec}.

In the later discussion, we consider the estimated 90\% upper-limit, \lumx\ $<3.0\times10^{43}$\,\lumcgs 
with $\Gamma$\,=\,$1.5$ and \nh\,=\,$1.5\times10^{24}$\,\pscm,
as a fiducial value unless otherwise stated\footnote{We also estimate the upper-limit using Xspec fit with C-statistic. The upper-limits from the two methods are consistent within a factor of 1.1.}.
The X-ray studies of ULIRGs \citep{Teng2014, Teng2015, Iwasawa2017, Iwasawa2018, Oda2017, Tombesi2017, Xu2017, Toba2020} show that \nh\ varies in a range from $10^{23}$\,\pscm\ (e.g., IRAS F05189–2524) to $3\times10^{24}$\,\pscm\ (e.g., IRAS F13120–5453), 
hence \nh\,=\,$1.5\times10^{24}$\,\pscm\ is a reasonable assumption to account for Compton-thick obscuration. 

Using the empirical relation of \citet{Lehmer2010}, the estimated \lumx\ contributed by high-mass X-ray binaries (HMXBs) in the host galaxy is 
$\sim$\,2$\times10^{42}$\,\lumcgs\ with stellar mass of $10^{11}$\,\msun\ and star formation rate of 1000\,\sfrunit, which is much lower than the 90\% upper-limit \lumx. 
Therefore hereafter we ignore the contribution of HMXBs and consider the 90\% upper-limit \lumx\ as the upper-bound of AGN X-ray luminosity. 

% Different choices of the background-extraction region  (e.g., a circular region in the
% source vicinity but outside the 120 radius aperture) do not appear to affect the background estimate systematically, 
% and the number of background counts generally fluctuates at the 􏰁10\% level. 

%%%%%%%%%%%%%%%%%%%%%%%%%%%%%%%%%%%%%%%%%%%%%%%%%%%%%%%%%%%%%%%%%%%%%%%%%%%%%%%%
%%%%%%%%%%%%%%%%%%%%%%%%%%%%%%%%%%%%%%%%%%%%%%%%%%%%%%%%%%%%%%%%%%%%%%%%%%%%%%%%

\begin{figure*}[ht!]
    \begin{center}
    \hspace{-38pt}
	\includegraphics[width=0.45\textwidth]{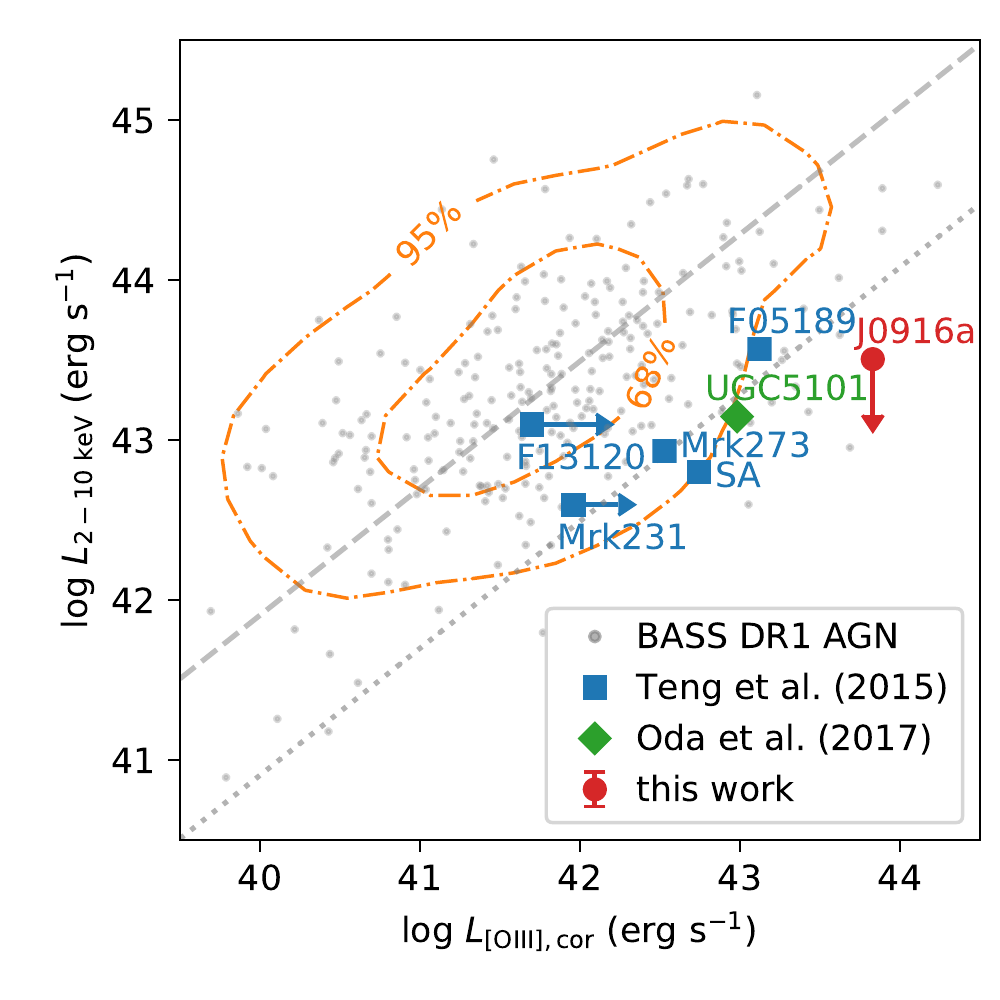}
	\includegraphics[width=0.45\textwidth]{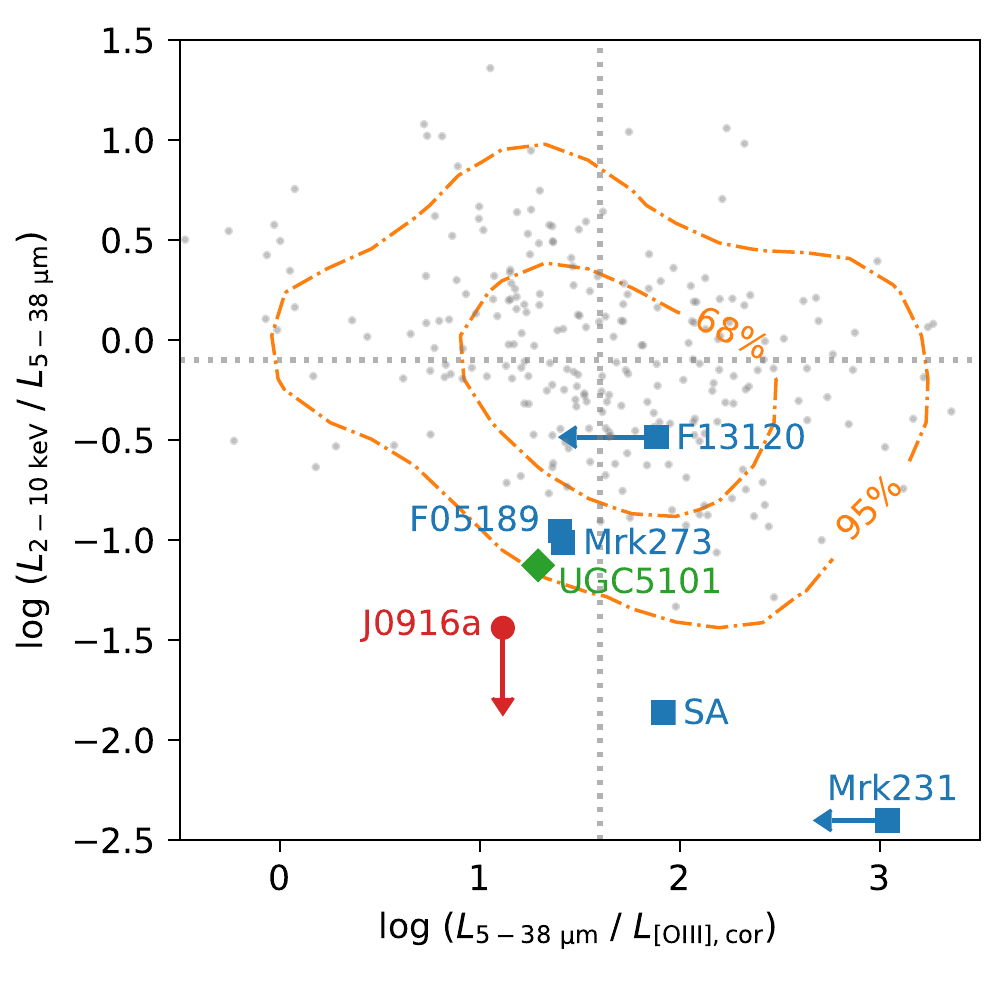}
    \end{center}
    \vspace{-25pt}
	\caption{
	\textbf{Left:} 
	Intrinsic \lumx\ vs. corrected $L_{\rm [OIII]}$ of J0916a compared to ULIRGs \citep{Teng2015, Oda2017} and normal AGNs from the BASS DR1 sample \citep{Koss2017}. 
	The red circle and downward arrow show the upper-limit \lumx\ of J0916a estimated assuming Compton-thick (\nh\,=\,$1.5\times10^{24}$\,\pscm) and Compton-thin obscuration ($10^{23}$\,\pscm), respectively. 
	The blue square and rightward arrow of F13120 denote the converted $L_{\rm [OIII]}$ from $L_{\rm [OIV]}$ using an empirical ratio of Seyfert 2 galaxies \citep{LaMassa2010} and an average $L_{\rm [OIII]}/L_{\rm [OIV]}$ ratio of the other ULIRGs, respectively. 
	The blue square and rightward arrow of Mrk\,231 denote the uncorrected and corrected $L_{\rm [OIII]}$ with a typical dust extinction of Seyfert 1 galaxies \citep{LaMassa2010}, respectively. 
	The orange contours show the 68\% ($1\sigma$) and 95\% ($2\sigma$) distribution ranges of the BASS DR1 AGNs (grey dots). 
	The grey dashed and dotted lines show the empirical \lumx-$L_{\rm [OIII]}$ relation from \citet{Ueda2015} and the $-1$ dex position under the relation, respectively. 
	\textbf{Right:} 
	\lumx/\lummir\ vs. \lummir/$L_{\rm [OIII]}$. 
	A small $y$-value denotes the corona ($<$\,10\,pc) fading compared to torus ($\sim$\,10\,pc), 
	while a small $x$-value denotes the torus fading compared to NLR (1--10 kpc). 
	The horizontal and vertical dotted lines show the median values of BASS DR1 AGNs. 
	Other legends are the same as in the left panel. 
	J0916a is the most extreme ULIRG with both fading corona and torus relative to the normal AGN sample. 
	}
	\label{fig:LX_LMIR_LOIII}
\end{figure*}

%%%%%%%%%%%%%%%%%%%%%%%%%%%%%%%%%%%%%%%%%%%%%%%%%%%%%%%%%%%%%%%%%%%%%%%%%%%%%%%%
%%%%%%%%%%%%%%%%%%%%%%%%%%%%%%%%%%%%%%%%%%%%%%%%%%%%%%%%%%%%%%%%%%%%%%%%%%%%%%%%
%%%%%%%%%%%%%%%%%%%%%%%%%%%%%%%%%%%%%%%%%%%%%%%%%%%%%%%%%%%%%%%%%%%%%%%%%%%%%%%%

\section{Discussions} \label{sec:discussions}
\subsection{X-ray faintness relative to NLR emission} \label{sec:discuss_X_NLR}

Figure \ref{fig:LX_LMIR_LOIII} (left panel) shows the intrinsic \lumx\ and extinction corrected $L_{\rm [OIII]}$ of J0916a compared to six well-studied nearby ($z<0.1$) ULIRGs from \citet{Teng2014, Teng2015} and \citet{Oda2017}. 
Four of the referred ULIRGs show Compton-thin obscuration, i.e., IRAS F05189–2524 (hereafter F05189), 
Mrk\,231\footnote{We employ the intrinsic \lumx\ from MYTorus model fitting in \citet{Teng2014}.}, 
Mrk\,273, and 
Superantennae\footnote{Since the reported \lumx\ of SA in \citet{Teng2015} was estimated without intrinsic obscuration, 
we replace with the result of \citet{Brightman2011} for SA.} 
(IRAS F19254–7245, hereafter SA);
the rest two ULIRGs host a Compton-thick level absorber, i.e., IRAS F13120–5453 (hereafter F13120) and 
UGC\,5101\footnote{We employ the intrinsic \lumx\ from ``Model II'' fitting in \citet{Oda2017}.}.
The corrected $L_{\rm [OIII]}$ of the referred ULIRGs are collected from 
\citet[][F05189 and Mrk\,273]{Veilleux1999}, \citet[][SA]{Buchanan2006}, \citet[][UGC\,5101]{Moustakas2006}, and 
\citet[][Mrk\,231]{Singh2011}\footnote{Since the Balmer decrement is not available, the $L_{\rm [OIII]}$ of Mrk\,231 is not corrected and shown as lower-limit in Figure \ref{fig:LX_LMIR_LOIII}.}. 
The \oiii\ detection is not available for F13120 and we employ the $L_{\rm [OIII]}$ converted from $L_{\rm [OIV]}$ %\citep{Fernandez2016} 
with the typical ratio of Seyfert 2 galaxies\footnote{
Note that the converted $L_{\rm [OIII]}$ of F13120 could be underestimated, 
since the average $L_{\rm [OIII]}/L_{\rm [OIV]}$ ratio of the four ULIRGs (F05189, Mrk\,273, SA, UGC\,5101), $\sim$\,1.0\,dex, 
seems higher than the ratio of Seyfert galaxies \citep[0.59\,dex,][]{LaMassa2010}.
}\,\citep{LaMassa2010}.
The contribution to $L_{\rm [OIII]}$ from star formation in the ULIRGs is estimated using the empirical relation of starbursts \citep{Gurkan2015}, 
which is only 0.5\%--2\% of the total $L_{\rm [OIII]}$ and thus, 
we ignore the star formation contamination and consider that the $L_{\rm [OIII]}$ fully accounts for AGN activity. 

In order to compare J0916a and other ULIRGs to normal AGNs, we also include the catalog of the Swift-BAT AGN Spectroscopic Survey \citep[BASS DR1,][]{Koss2017}. 
In total, 243 AGNs are selected with S\,/\,N\,$>$\,3 for \ha, \hb, and \oiii\ emission lines, 
which consists of 135 Seyfert 1 galaxies (including Seyfert 1.2 to 1.9) and 108 Seyfert 2 galaxies. 
The corrected $L_{\rm [OIII]}$ and \lumx, which are converted from the Swift 14--195 keV luminosity and empirical ratio of \citet{Ricci2017}, 
are shown in the left panel of Figure \ref{fig:LX_LMIR_LOIII} with 68\% ($1\sigma$) and 95\% ($2\sigma$) distribution contours. 

The ratio, \lumx\,$/\,L_{\rm [OIII]}$, can be considered 
as an indicator of the current X-ray faintness in the nuclear region relative to the past AGN activity. 
All of the ULIRGs (except for F13120) in Figure \ref{fig:LX_LMIR_LOIII} (left panel) show lower \lumx\,$/\,L_{\rm [OIII]}$ compared to the normal AGNs. 
If we adopt a modified faintness ratio, $f_{\rm X}$\,=\,$L_{\rm 2-10\,keV}\,/\,L_{\rm X,\,[OIII]}$, 
where $L_{\rm X,\,[OIII]}$ is the 2--10 keV luminosity converted from $L_{\rm [OIII]}$ using the empirical relation of normal AGNs \citep{Ueda2015}, 
then the referred ULIRGs (except for F13120) possess an average $f_{\rm X}$ of 10\%, 
implying a general trend of X-ray deficit in the ULIRGs. 
J0916a shows $f_{\rm X}$\,$<$\,3.6\% assuming \nh\,=\,$1.5\times10^{24}$\,\pscm, 
which is even lower than the lowest $f_{\rm X}$\,$=$\,5.2\% of SA in the referred ULIRGs. 
The $f_{\rm X}$ of J0916a can be much lower ($<1.6\%$) if we consider Compton-thin obscuration (\nh\,=\,$10^{23}$\,\pscm). 
The extremely low $f_{\rm X}$ of J0916a suggests a most extreme case of X-ray declining among the ULIRGs, which is possibly related to a strong nuclear wind implied by its highest $\dot{E}_{\rm k}$ ionized outflow in the galaxy scale (see Section \ref{sec:discuss_feedback} for a detailed discussion).

%%%%%%%%%%%%%%%%%%%%%%%%%%%%%%%%%%%%%%%%%%%%%%%%%%%%%%%%%%%%%%%%%%%%%%%%%%%%%%%%

\subsection{Fading AGN central engine indicated by both faint corona and torus radiation} \label{sec:discuss_fading}

The MIR emission from the dusty torus can be used as an intermediate indicator of AGN activity with its spatial scale ($\sim$\,10\,pc) between 
X-ray emitting corona \citep[$<$\,0.1\,pc,][]{Dai2010} and NLR (1--10 kpc), 
and it traces AGN activity in the last several 10 years \citep[e.g.,][]{Ichikawa2017b}. 
%In addition to X-ray radiation coming from the central region \citep[$\sim10\ r_{\rm g}$,][]{Dai2010}, 
%The MIR (5--38\,\micron) luminosity of the AGN in J0916a is $L_{\rm MIR}=8.8\times10^{44}$\,\lumcgs, which is integrated with the best-fit AGN SED including the torus thermal and scattering radiation as well as the transmitted primary emission \citep{Chen2019, Chen2020}. 
In order to make a fair comparison to J0916a, we collect the optical-IR photometries from 
the Sloan Digital Sky Survey (SDSS), 
the Two Micron All Sky Survey (2MASS), 
the Wide-field Infrared Survey Explorer (WISE), 
the Spitzer Space Telescope, 
the Infrared Astronomical Satellite (IRAS), 
and AKARI for the six referred ULIRGs, 
%using the NASA/IPAC Extragalactic Database (NED), 
and then perform SED fitting with \texttt{CIGALE} \citep{Boquien2019} following the configuration of \citet{Chen2019}.
%Noll2009, and \citet{Fritz2006} torus model
The $L_{\rm 5-38\,\mu m}$ of the referred ULIRGs are integrated with the best-fit AGN SEDs including the torus thermal and scattering radiation as well as the transmitted primary emission.
The $L_{\rm 5-38\,\mu m}$ of the BASS AGN sample are adopted from \citet{Ichikawa2019s}. 

The right panel of Figure \ref{fig:LX_LMIR_LOIII} shows the values of two ratios, i.e., \lumx/\lummir\ and \lummir/$L_{\rm [OIII]}$. 
A low \lumx/\lummir\ relative to normal AGNs suggests AGN fading in the core region (0.1\,pc compared to 10\,pc), 
while a low \lummir/$L_{\rm [OIII]}$ denotes AGN fading in the circumnuclear region (10\,pc compared to several kpc). 
With this diagram the ULIRGs can be separated into three groups:
1) F13120 shows the properties similar to normal AGNs;
2) SA and Mrk\,231 possess decayed coronae but their tori are still luminous;
3) the other four ULIRGs including J0916a show both faint corona and torus radiation. 
However, the fading ratios of F05189, Mrk\,273, and UGC\,5101 are moderate and within the 95\% ($2\sigma$) distribution range of the BASS AGN sample. 
J0916a locates at a unique position out of the 95\% range of the BASS sample, which indicates that 
both of its core (corona) and circumnuclear (torus) AGN emission are drastic declining. 

%%%%%%%%%%%%%%%%%%%%%%%%%%%%%%%%%%%%%%%%%%%%%%%%%%%%%%%%%%%%%%%%%%%%%%%%%%%%%%%%

\subsection{Powerful outflow with fading central engine: the cumulative effect of AGN feedback can be limited} \label{sec:discuss_feedback}

A possible scenario to explain both of the faintness in corona (X-ray) and torus (MIR) in J0916a is that 
the primary radiation from the AGN accretion disk is currently in a fading status. 
Suggested by the connection between the multi-phase outflows observed in nearby ULIRGs such as Mrk\,231 \citep{Feruglio2015} and F05189 \citep{Smith2019}, 
the powerful galaxy-scale ionized outflow in J0916a could imply an ultrafast nuclear wind
when the \oiii\ outflow was launched with 
a super-Eddington accretion.
The Eddington ratio ($\lambda_{\rm Edd}$) is estimated to be 5.7, with the bolometric luminosity converted from $L_{\rm [OIII]}$ 
and the mass of SMBH ($2.7\times10^7$\,\msun) estimated using the stellar mass of the host galaxy\footnote{To be compared, the $\lambda_{\rm Edd}$ estimated from $L_{\rm bol,\,torus}$ and $L_{\rm bol,\,corona}$ are 1.4 and $<0.3$, respectively.}.
The nuclear wind could blow out the surrounding gas and dust, weaken the development of corona and torus, and suppress the fueling to the SMBH. 

With the extent of the \oiii\ outflow (4\,kpc) and the maximum outflow velocity ($\sim$\,2000\,\kms), 
the fading process could happen in an outflow-traveling timescale of $\sim$\,2\,Myr, or a light-traveling timescale of $\sim$\,$10^4$ years. 
Namely, the fading timescales are much shorter than the duration of starburst in ULIRGs, i.e., $\sim$\,100\,Myr \citep[e.g.,][]{Hopkins2008}, 
implying that 
although the AGN can drive an powerful, fast wind in the galaxy, 
e.g., the outflow velocity of J0916a even exceeds the escape velocity of the host halo \citep{Chen2020}, 
it can be decayed in a relatively short period, 
and the cumulative effect of AGN feedback on the stellar build-up in the galaxy can be limited 
as indicated by the high star formation rate of 1000\,\sfrunit.
%Both quenching of AGN activity \citep[e.g.,][]{Ichikawa2019b} and violent variability \citep[e.g.,][]{Miniutti2012} can be one of possible mechanisms of the fading AGN scenario of J0916a.

%%%%%%%%%%%%%%%%%%%%%%%%%%%%%%%%%%%%%%%%%%%%%%%%%%%%%%%%%%%%%%%%%%%%%%%%%%%%%%%%

\subsection{Other possibilities and future work} \label{subsec:discuss_other}

In the above discussion a Compton-thick level absorber, \nh\,=\,$1.5\times10^{24}$\,\pscm, is assumed, 
which accounts for the heavy obscuration in ULIRGs \citep[e.g.,][]{Teng2015}. 
However, it is hard to rule out the possibility of a very Compton-thick absorber, e.g., \nh\,$\sim$\,$10^{25}$\,\pscm, locating on the line of sight and shielding the X-ray emitting corona. 
Significant constraint on the nuclear gas column density in J0916a is over the capability of the current X-ray instruments due to its faintness at a relatively high redshift. 
Sub-millimeter (submm) observations of dust emission with a high spatial resolution, e.g., ALMA, provide an alternative method to study such heavily obscured environment in galaxies \citep[e.g.,][]{Scoville2015}. 
Note that although a very Compton-thick absorber may explain the observed X-ray faintness in J0916a, 
we still need other scenarios for the explanation on the MIR faintness of the torus.  

%Recent studies show that the X-ray deficit (e.g., low $L_{\rm X}/L_{\rm MIR}$ and $L_{\rm X}/L_{\rm UV}$) is associated to high $\lambda_{\rm Edd}$, possibly as a result of the weakening of corona when the inner edge of accretion disk moves inwards \citep[e.g.,][]{Toba2019}. 
Recent studies showed that the weakness of X-ray emission is associated to high $\lambda_{\rm Edd}$ \citep[e.g.,][]{Toba2019} and fast disk winds \citep[e.g.,][]{Zappacosta2020}, in which the X-ray deficit does not suggest the fading of the primary accretion disk radiation, but could be explained with the weakening of corona when the disk inner-edge moves inwards, or the cooling of corona due to the dense, so-called ``failed wind''.
Within those scenarios, the high $\lambda_{\rm Edd}$\,=\,5.7 of J0916a from its extended \oiii\ emission could reflect not only the past, but also the current AGN activity, i.e., the AGN is still active. 
However, the expected \lummir\ in this active scenario could be over three times brighter than the observed value, 
considering that the covering factor of torus keeps nearly constant with $\lambda_{\rm Edd}$\,$>$\,0.03 \citep{Ricci2017}.
With the observed low \lummir/$L_{\rm [OIII]}$, we tend to adopt the fading scenario as discussed in Section \ref{sec:discuss_feedback}, although it is difficult to rule out the active scenario due to the uncertainty in the estimation of NLR radiation.
In this work the \oiiiblong\ line is used as the indicator of AGN NLR radiation. 
Compared to the IR \oivlong\ emission line, \oiii\ emission is more likely contaminated by the ionized gas surrounding young O\,/\,B type stars, 
and is more sensitive to the dust extinction. 
We estimate the amount of star formation contamination in the total $L_{\rm [OIII]}$, which is only 0.6\% and can be ignored. 
However, the uncertainty in the extinction estimation due to the weakness of \hb\ line can results in a variation of 0.5 dex in the corrected $L_{\rm [OIII]}$. 
%The dust extinction was estimated to be $E(B-V)=1.0\pm0.3$ using Balmer decrement \citep{Chen2019}, which is close to the typical amount of dust attenuation in the local U/LIRGs \citep[e.g.,][]{Alonso2006, Garcia2009}. 
%The uncertainty in the extinction estimation can results in a variation of 0.5 dex in the corrected $L_{\rm [OIII]}$. 
High quality optical spectrum is required to reduce the uncertainty of extinction correction, which will be achieved with the awarded Gemini/GMOS IFU follow-up for J0916a. 
The IFU observation can also help us to better constrain the extent the size of the outflowing region. 
% concern on choice of NLR indicator. Possibly [OIV] < [OIII] in declined AGN.

In addition to the fading AGN scenario where the central AGN activity indicated by X-ray and MIR radiation is finally quenched, 
there is another possibility that the AGN in J0916a may brighten up again as shown in the ``changing-look AGNs'' (CLAGNs), 
which have strong variability by an order of magnitude and can rebrighten when their accretion states change from faint phase to bright phase in a timescale of 1--10 years \citep[e.g.,][]{Ricci2020}. %Ilic2020, 
If J0916a has a CLAGN and its X-ray emitting corona would recover \lumx\ of $\sim9\times10^{44}$\,\lumcgs\ 
as expected from the corrected $L_{\rm [OIII]}$, 
then the AGN can be significantly detected by NuSTAR, e.g., 
S\,/\,N\,$>$\,7 in 3--20 keV with an exposure of 100 ksec in a Compton-thick case (\nh\,$\sim$\,$10^{24}$\,\pscm). 
Therefore the future NuSTAR follow-up in a few years is necessary to test the CLAGN scenario. 
The submm observation with a high spatial resolution, e.g., ALMA, can be also helpful to constrain the CLAGN scenario.
Considering the spectral coverage of ALMA at $z$\,=\,0.49, CO\,J=4-3 line can be employed as a tracer of molecular gas in J0916a. 
The spatial resolution of ALMA can reach 0.01\arcsec\ (80\,pc) at Band 7 with the maximum baselines of 16.2 km, 
which provides the capability to determine the molecular gas reservoir in the circumnuclear region (100--200\,pc) of the AGN. 
If the ALMA observation reveals that the cold molecular gas has been cleaned up in the circumnuclear region,
the possibility of AGN rebrightening could be reduced without fueling to the SMBH. 

%%%%%%%%%%%%%%%%%%%%%%%%%%%%%%%%%%%%%%%%%%%%%%%%%%%%%%%%%%%%%%%%%%%%%%%%%%%%%%%%
%%%%%%%%%%%%%%%%%%%%%%%%%%%%%%%%%%%%%%%%%%%%%%%%%%%%%%%%%%%%%%%%%%%%%%%%%%%%%%%%

\acknowledgments

We thank the anonymous referee for the valuable comments and constructive suggestions. 
We thank Drs. Junxian Wang and Teng Liu for the fruitful discussion on mechanisms of X-ray faintness. 
This research has made use of the NuSTAR Data Analysis Software (NuSTARDAS) jointly developed by the ASI Space Science Data Center (SSDC, Italy) and the California Institute of Technology (Caltech, USA).


%\vspace{5mm}
\facilities{NuSTAR}


\software{NuSTARDAS, CIGALE}
%\vspace{-5mm}



%\appendix
%\section{Appendix information}

\bibliography{nustar_j0916a}{}
\bibliographystyle{aasjournal}



\end{document}